\documentclass[twocolumn,secnumarabic,amssymb, nobibnotes, aps, pra, superscriptaddress]{revtex4-1}
\usepackage{graphicx}
\usepackage{braket}
\usepackage{array}
\usepackage{amsmath}
\usepackage{here}
\usepackage{color}
\usepackage{ulem}
\usepackage{bm}

\begin{document}

\title{Coherent control of a local phonon in trapped ions using dynamical decoupling}%

\author{Ryutaro Ohira}%
\affiliation{Graduate School of Engineering Science, Osaka University, 1-3 Machikaneyama, Toyonaka, Osaka 560-8531, Japan}
\affiliation{Center for Quantum Information and Quantum Biology, International Advanced Research Institute, Osaka University, 1-3 Machikaneyama, Toyonaka, Osaka 560-8531, Japan}
\author{Shota Kume}%
\affiliation{Graduate School of Engineering Science, Osaka University, 1-3 Machikaneyama, Toyonaka, Osaka 560-8531, Japan}
\author{Kenji Toyoda}%
\email{toyoda@qiqb.osaka-u.ac.jp}
\affiliation{Center for Quantum Information and Quantum Biology, International Advanced Research Institute, Osaka University, 1-3 Machikaneyama, Toyonaka, Osaka 560-8531, Japan}

\date{\today}

\begin{abstract}
In this paper, we present a dynamical decoupling (DD) technique to coherently control the dynamics of a single local phonon in trapped ions. A 2$\pi$ rotation at a motional sideband transition flips the sign of the relevant local phonon state, resulting in cancellation of the phonon dynamics. In this work, we implement DD using single and multiple blue-sideband pulses to control a single local phonon in two $^{40}{\rm Ca}^{+}$ ions in a linear Paul trap. Our proposed DD technique can be used to engineer coupling between local phonon modes.
\end{abstract}


\maketitle

\section{Introduction}

Local phonons in trapped ions are considered bosonic systems consisting of individual harmonic oscillators coupled with each other. To date, phonon propagation has been experimentally observed in ions trapped in a double-well potential \cite{1,2,3}, in a harmonic potential of a linear Paul trap \cite{4,5,6,7,8,9,10,11}, and in a lattice potential of a two-dimensional trap \cite{12,13,14}.

A system of local phonons possesses certain advantages for implementing quantum computation and quantum simulation. For instance, local phonons are particularly applicable to Hubbard-type quantum simulations, such as the Bose--Hubbard model \cite{15,16}, the Jaynes--Cummings--Hubbard (JCH) model \cite{17,18,19,20,21,22}, and the Rabi--Hubbard model \cite{23}.

A local phonon system is analogous to photons propagating in a linear optical circuit \cite{24}. Probabilistic operations due to the absence of nonlinearity are likely to hinder the scale-up of photonic quantum systems. On the other hand, a local phonon system guarantees the deterministic generation of a single or multiple phonons and highly efficient detection of the motional state. In this context, scalable boson sampling with local phonons in trapped ions \cite{25, 26} and continuous-variable (CV) quantum computation \cite{27} have been proposed.

An essential concept in these applications is a phonon circuit or a local-phonon-based quantum simulator (Fig.\,1), analogous to a photonic circuit. A local-phonon-based quantum simulator consists of the following three components: (1) state preparation, (2) interaction between different phonon modes, and (3) detection of the output state. The first step is state preparation, where the motional state can be initialized using ground-state cooling \cite{28} and engineered in many ways \cite{29, 30, 31}. After preparation, each local phonon mode experiences unitary evolution, and the output state can then be detected. To date, projective measurement for a single ion \cite{31,32,33,34,35} or multiple ions \cite{9} has been developed.

\begin{figure}[t]
\centering
  \includegraphics[width=8.5cm]{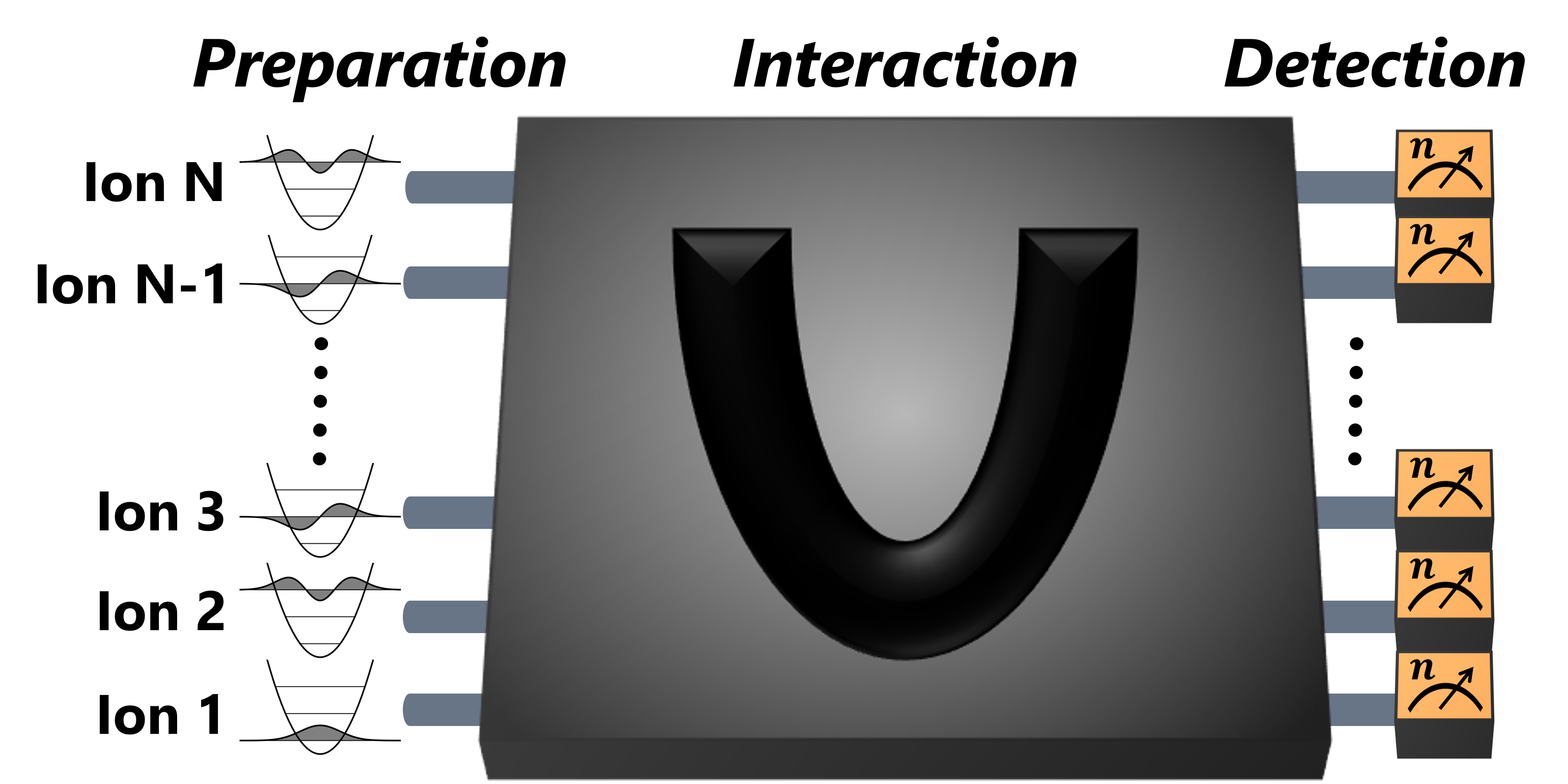}
\caption{\label{fig1} Local-phonon-based quantum simulator (phonon circuit). The three steps of the process are shown: (1) motional state preparation, (2) unitary evolution, and (3) detection of the phonon number distribution of the output state.}
\end{figure}

An advantage of using local phonon modes in an ion string, instead of collective motional modes, is the straightforward scalability. The local phonon modes are a collection of radial motional modes associated with each ion \cite{15}. Increasing the number of ions in a string results in spectral crowding of the collective motional modes. While it is necessary to combine the addressing of particular ions with precise tuning in the frequency domain for the efficient addressing of a particular collective motional mode, local phonon modes can be addressed separately by illuminating the corresponding ion with a laser beam. 

Preparing and measuring states in local phonon modes have been demonstrated as relatively direct extensions of the case for a single motional mode. What is important and still needs to be explored is tuning the couplings between local phonon modes in a versatile manner. 

A few attempts have been made to control the quantum dynamics of local phonons using optical pulses. A phonon blockade \cite{8,11} is an example. The violation of energy conservation due to off-resonantly coupled oscillators prevents the individual bosons from hopping to other ion sites.

In this work, we present the dynamical decoupling (DD) \cite{36,37} of local phonon modes coupled to an internal degree of freedom. Our work is inspired by a method proposed in Ref.\,\cite{25}. In the method in Ref.\,\cite{25}, local phonons evolve under the Hamiltonian $\hat{H}$ for a period of $t$. An instantaneous off-resonant motional sideband pulse is then applied, inducing a $\pi$ phase shift to a particular local phonon mode. Accordingly, the local phonons evolve under the Hamiltonian $-\hat{H}$ for the next period of $t$, being subject to a cancellation of the dynamics for the first and second half periods. However, if practical experimental parameters are considered, the possible interaction rate of the phase-shift operation is comparable or slower than the typical hopping rate.

Here, as a step toward the full implementation of DD as proposed in Ref.\,\cite{25}, we demonstrate the coherent control of a local phonon using optical pulses, and thereby realize DD of two local phonon modes in the presence of a single local phonon. The ions are driven with resonant motional sideband pulses instead of off-resonant pulses. The resonant excitation allows for the realization of fast local phonon manipulation. In the experiments, we manipulate a single local phonon. Since the number of ions does not limit our DD technique, the scheme can be applied to a long ion chain. DD techniques have previously been applied to spin or qubit systems \cite{36, 37}. To our knowledge, however, DD of phonon modes in trapped ions has not yet been realized. The DD technique demonstrated here can be applied to manipulating local phonons in trapped ions.

\section{Dynamical decoupling of local phonons in trapped ions}

\subsection{General idea}

Here, we describe the general idea of DD of local phonon modes in trapped ions. We consider local phonon dynamics in a two-ion chain to explain the scheme, but the scheme can be applied to an $N$-ion chain. 

In the interaction picture, the Hamiltonian for the local phonon mode along the radial direction is written as follows:
\begin{equation}
\hat{H}_{\rm Hop} = \frac{\kappa_{12}}{2}(\hat{a}_{1}\hat{a}_{2}^{\dagger}+\hat{a}_{1}^{\dagger}\hat{a}_{2}),
\end{equation}
where $\kappa_{12}$ is the hopping rate between Ion 1 and Ion 2. $\hat{a}_{i}^{\dagger}$ and $\hat{a}_{i}$ are the creation and annihilation operators of the local phonon mode along the radial direction of the $\it{i}$th ion $(i=1,2)$.

The basic idea of the DD method is as follows. First, we prepare two ions in a quantum state: 
\begin{equation}
\ket{\psi(0)} = \ket{\psi_1}\otimes\ket{\psi_2},
\end{equation}
where
\begin{equation}
\ket{\psi_{1(2)}} = \sum_{n=0}^{N}c_{n,1(2)}\ket{n}.
\end{equation}
Here, $c_{n,1}$ and $c_{n,2}$ are the probability amplitudes of $\ket{n}$ of Ion1 and Ion2, satisfying $\sum_{n=0}^{N}|c_{n,1(2)}|^2=1$.

Let the system evolve under the Hamiltonian $\hat{H}_{\rm Hop}$ for a period $t$:
\begin{equation}
\ket{\psi(t)} = e^{-i\hat{H}_{\rm Hop}t}\ket{\psi(0)}.
\end{equation}

Then, a phase-shift operation is applied to a particular local phonon mode (here, it is assumed to be the $k$th mode, where $k=1,2$):
\begin{equation}
\hat{U} = e^{i\theta_{k}\hat{a}_{k}^{\dagger}\hat{a}_{k}},
\end{equation}
where $\theta_{k}$ is the phase shift. This operator transforms $\hat{a}_{k}^{\dagger}$ and $\hat{a}_{k}$ as 
\begin{equation}
\hat{U}^{\dagger}\hat{a}_{k}^{\dagger}\hat{U}=\hat{a}_{k}^{\dagger}e^{-i\theta_k},
\end{equation} 
\begin{equation}
\hat{U}^{\dagger}\hat{a}_{k}\hat{U}=\hat{a}_{k}e^{i\theta_k}, 
\end{equation} 
respectively \cite{40}.

When $\theta_{k}=\pi$, the transformed operators are $-\hat{a}_{k}^{\dagger}$ and $-\hat{a}_{k}$. Accordingly, the sign of the Hamiltonian is flipped. Therefore, the hopping dynamics is ``time-reversed'' during the next period of $t$:
\begin{equation}
\ket{\psi(2t)} = e^{i\hat{H}_{\rm Hop}t}e^{-i\hat{H}_{\rm Hop}t}\ket{\psi(0)}.
\end{equation}

\subsection{Physical implementation of dynamical decoupling}

Here, we describe how we can implement the DD of local phonon modes in trapped ions.

\subsubsection{Dynamical decoupling based on a dispersive Jaynes--Cummings interaction}

First, we briefly describe the scheme proposed by Shen $et$ $al$.\,\cite{25} as a reference for comparison against our scheme. The phase shift of the local phonon mode is induced by the dispersive Jaynes--Cummings (JC) interaction between the ions with an off-resonant sideband pulse. It is assumed that the two internal states $\ket{\downarrow}$ and $\ket{\uparrow}$ are used. When the $k$th ion is excited by a resonant red-sideband pulse, the resulting Hamiltonian for the ion is as follows \cite{38}:
\begin{equation}
\hat{H}_{\rm{RSB}} = g_r(\hat{a}_{k}^{\dagger}\hat{\sigma}^{-}+\hat{a}_{k}\hat{\sigma}^{+}),
\end{equation}
where the raising and lowering operators for the $k$th ion are defined as $\hat{\sigma}^{+}=\ket{\uparrow}\bra{\downarrow}$ and $\hat{\sigma}^{-}=\ket{\downarrow}\bra{\uparrow}$. $2g_r$ is the Rabi frequency at the red-sideband transition.

Now, an ion in a two-ion chain is driven with an off-resonant red-sideband pulse whose detuning from the sideband transition is $\Delta$. For $\Delta\gg 2g_{r}\sqrt{n}$, where $n$ is the phonon number, this off-resonant excitation results in a dispersive interaction between the phonon mode and the internal states \cite{39}:
\begin{equation}
\hat{H} = \chi\hat{\sigma}_{z}\hat{a}_{k}^{\dagger}\hat{a}_{k},
\end{equation}
where $\chi=g_{r}^2/\Delta$ and $\sigma_{z}=\ket{\uparrow}\bra{\uparrow}-\ket{\downarrow}\bra{\downarrow}$. Using this interaction, a phase-shift operation on the local phonon mode is realized:
\begin{equation}
\hat{U} = e^{i\theta_{i}\hat{a}_{k}^{\dagger}\hat{a}_{k}}.
\end{equation}
Here, $\theta_{k}=\chi{T}$ is the phonon number dependent phase shift, where $T$ is the pulse duration.

\begin{figure}[t]
\centering
  \includegraphics[width=7.0cm]{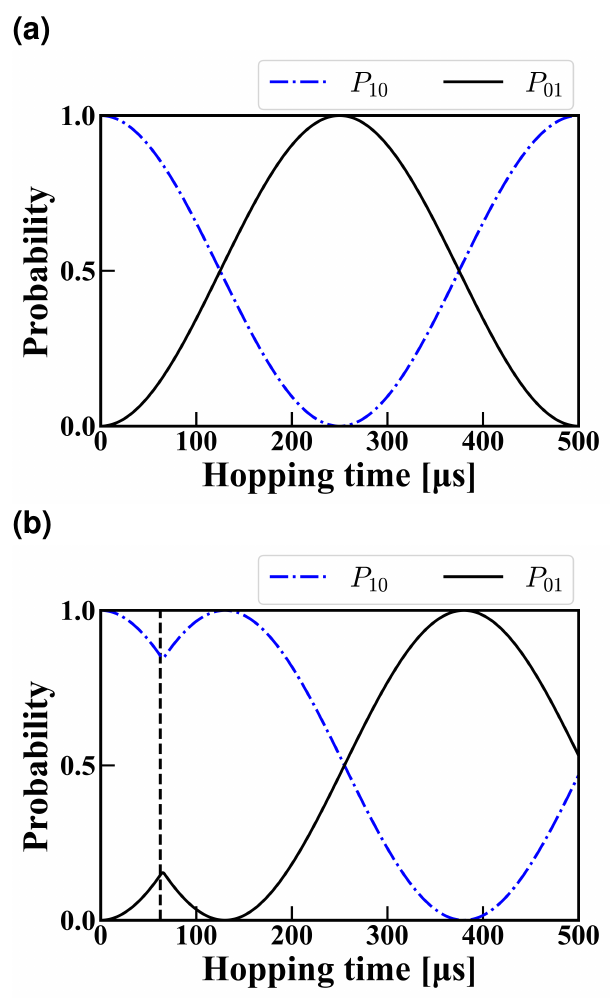}
\caption{\label{fig1} Single-local-phonon dynamics with and without a $\pi$ phase shift. (a) Free hopping of a single local phonon. (b) Single phonon dynamics with a phase-shift operation with the parameters of $\chi/\kappa_{12}=$50. In the simulation, the phase-shift operator is applied to the second ion at $t=\pi/4\kappa_{12}=62.5$ $\mu s$ (vertical dashed line).}
\end{figure}

The effect of a phase-shift operation on single-local-phonon propagation is shown in Fig.\,2. Here, we employ the Liouville equation for the density matrix with the parameter $\kappa_{12}/{\rm 2}\pi$ = 2 kHz. As an initial state, the quantum states of the ions are prepared in $\ket{\psi_{\rm Init}}=\ket{\downarrow_{1},1}\otimes\ket{\downarrow_{2},0}\equiv\ket{1,0}$. For simplicity, no decoherence process is incorporated in the results of Fig.\,2. The blue dot-dashed and black solid curves represent the probability of finding $\ket{1,0}$ ($P_{10}$) and $\ket{0,1}$ ($P_{01}$), respectively. A numerically calculated result of free hopping is shown in Fig.\,2(a). The phase-shift effect on single-local-phonon propagation is shown in Fig.\,2(b). The parameter $\chi/\kappa_{12}=$50 is used in Fig.\,2(b). In the simulation, the phase-shift operation is applied to the second ion at $t=\pi/4\kappa_{12}=$62.5 $\mu s$.

The dynamics of multiple local phonons can also be controlled with the phase-shift operation. Here, we again employ the Liouville equation for the density matrix with the parameter $\kappa_{12}/{\rm 2}\pi$ = 2 kHz. As an initial state, the quantum states of the ions are prepared in $\ket{\psi_{\rm Init}}=\ket{2,0}$. In Fig.\,3, the blue dot-dashed, black solid, and red dashed curves represent the probability of finding $\ket{2,0}$ ($P_{20}$), $\ket{1,1}$ ($P_{11}$), and $\ket{0,2}$ ($P_{02}$), respectively. A numerically calculated result of free hopping is shown in Fig.\,3(a). The phase-shift effect on two-local-phonon propagation is shown in Fig.\,3(b) using the parameter $\chi/\kappa_{12}=$50. In the simulation, the phase-shift operation is applied to the second ion at $t=\pi/4\kappa_{12}=62.5$ $\mu s$.

\begin{figure}[t]
\centering
  \includegraphics[width=7.0cm]{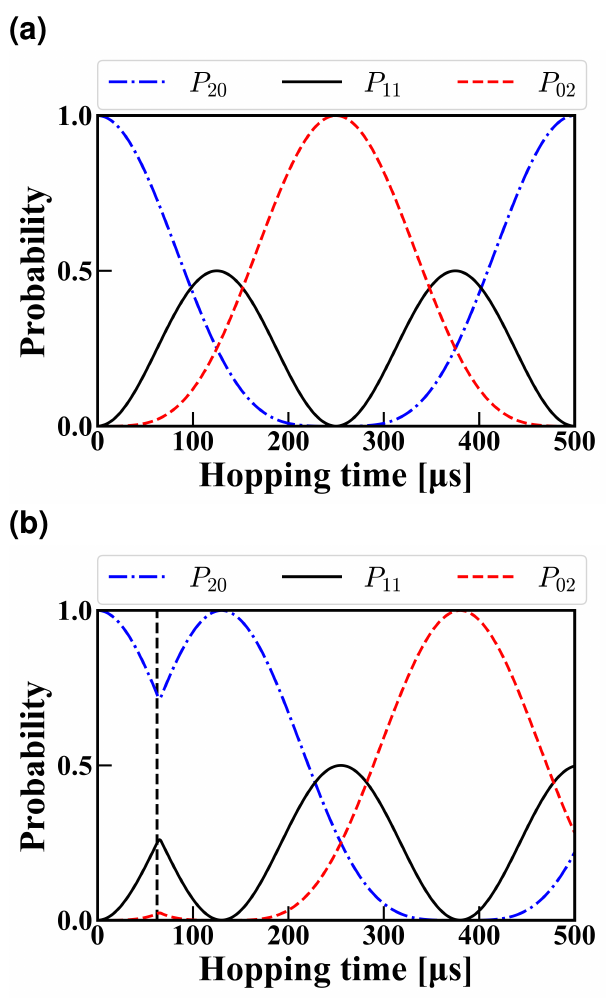}
\caption{\label{fig1} Two-local-phonon dynamics with and without a $\pi$ phase shift. (a) Free hopping of two local phonons. (b) Two-phonon dynamics with a phase-shift operation with the parameters of $\chi/\kappa_{12}=$50. In the simulation, the phase-shift operator is applied to the second ion at $t=\pi/4\kappa_{12}=62.5$ $\mu s$ (vertical dashed line).}
\end{figure}

In this way, DD of local phonons can be realized. However, in reality, phase-shift operations may not be applied almost instantly, as assumed above, but may take a non-negligible time. This time is sufficiently short if $\chi\gg\kappa_{12}$. The typical values for the parameters obtained with the present conditions of our experiments ($\chi/2\pi\sim0.50$--$1.25$ kHz and $\kappa_{12}/2\pi\sim1.0$--$10$ kHz, where $2g_{r}/2\pi\sim20$--$50$ kHz and $\Delta=10\times2g_{r}$ are used) do not satisfy this condition. Therefore, it is not realistic in the current conditions to implement the phase-shift operation based on the dispersive interaction.

\subsubsection{Dynamical decoupling based on resonant sideband pulses}

In our scheme, we employ a resonant sideband pulse instead of an off-resonant pulse. We assume that only the motional Fock states up to the first excited state ($\ket{n=1}$) are populated, and the internal state is in $\ket{\downarrow}$. If a red sideband pulse is applied to an ion, the probability amplitude of $\ket{\downarrow,1}$ undergoes a Rabi cycle, while that of $\ket{\downarrow, 0}$ remains unchanged. If the length of the red sideband pulse is adjusted so that the probability amplitude of $\ket{\downarrow,1}$ completes a Rabi cycle via $\ket{\uparrow, 0}$ (2$\pi$ rotation in that specific red sideband transition), its phase is changed by $\pi$. This phase change can be interpreted as a geometric phase acquired in the resonant Rabi cycle within the two-level system $\{\ket{\downarrow,1},\ket{\uparrow,0}\}$. In contrast, the probability amplitude of $\ket{\downarrow, 0}$ undergoes no phase change. The overall effect of the red sideband pulse is equivalent to applying $\hat{U}$ in Eq.\ (5) with $\theta_k=\pi$ provided that the initial state is limited to within the manifold spanned by $\{\ket{\downarrow, 0},\ket{\downarrow, 1}\}$, thus enabling DD involving similar time-reversed dynamics.

The clear advantage of this scheme is that the use of a resonant sideband pulse leads to a faster implementation of DD compared with the case when dispersive phase shifts are used. The disadvantage is that it is not applicable to motional Fock states with a quantum number higher than 1. We could avoid this disadvantage by limiting the application of this scheme to initially unoccupied motional modes.

\begin{figure}[t]
\centering
  \includegraphics[width=7.0cm]{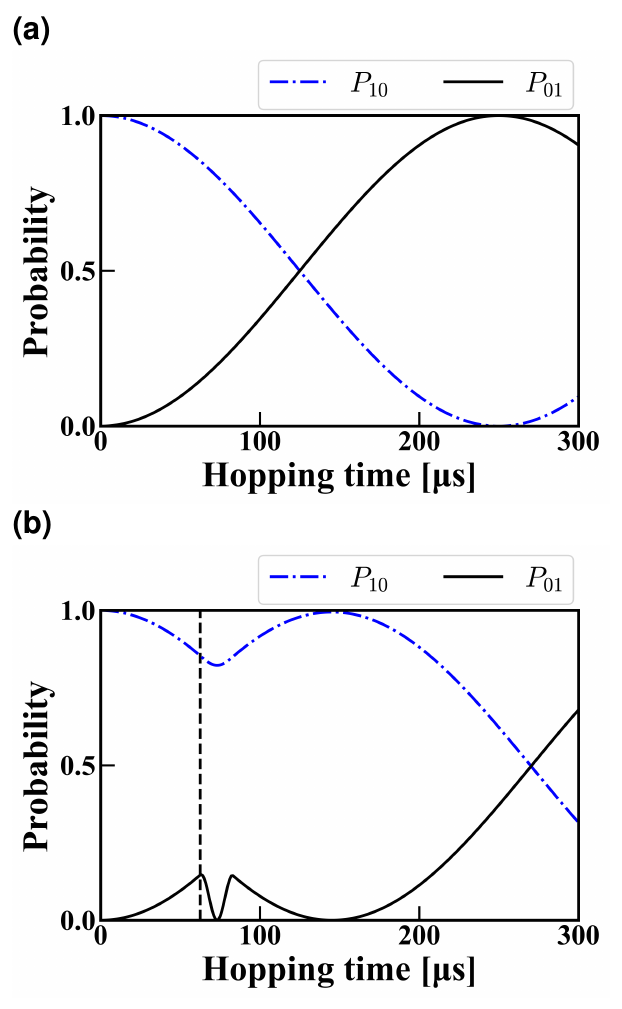}
\caption{\label{fig1} Single-local-phonon dynamics with and without a 2$\pi$ red-sideband pulse. (a) Free hopping of a single local phonon. (b) Single-phonon dynamics with a 2$\pi$ red-sideband pulse with the parameter $2g_{\rm r}/\kappa_{12}=$25. In the simulation, a 2$\pi$ red-sideband pulse is applied to the second ion at $t=\pi/4\kappa_{12}=62.5$ $\mu s$ (vertical dashed line).}
\end{figure}

The rotation operation for the red-sideband transition for the $k$th ion is expressed as follows:
\begin{equation}
\hat{R}(\theta, \phi) = \exp\Biggl[i\frac{\it{\theta}}{\rm 2}(e^{i\phi}\hat{a}_{k}^{\dagger}\hat{\sigma}^{-}+e^{-i\phi}\hat{a}_{k}\hat{\sigma}^{+})\Biggr], 
\end{equation}
where $\theta$ and $\phi$ denote the angle of the qubit rotation and the azimuth angle of the rotation axis, respectively. Assuming that the state of the ion is $\ket{\downarrow,1}$, the state evolves into $-\ket{\downarrow,1}(=\hat{R}(2\pi,\phi)\ket{\downarrow,1})$ after a $2\pi$ rotation at the red-sideband transition (in fact, the pulse area depends on the initial motional Fock state; here and hereafter, the first red sideband transition, $\{\ket{\downarrow,1},\ket{\uparrow,0}\}$, or the first blue sideband transition, $\{\ket{\downarrow,0},\ket{\uparrow,1}\}$, is taken as the reference transition for determining the pulse area of a sideband pulse). Note that the discussion here also applies to the blue-sideband interaction provided that the two internal states $\ket{\downarrow}$ and $\ket{\uparrow}$ are swapped. In the present study, we use the blue-sideband interaction for the local phonon manipulation instead of the red-sideband interaction because it provides better phonon manipulation fidelity for a technical reason. 

Single-local-phonon propagation with and without a 2$\pi$ red-sideband pulse is shown in Fig.\,4. As an initial state, the quantum states of the ions are prepared in $\ket{\psi_{\rm Init}}\equiv\ket{1,0}$. The blue dot-dashed and black solid curves represent the probability of finding $\ket{1,0}$ ($P_{10}$) and $\ket{0,1}$ ($P_{01}$), respectively. A numerically calculated result of free hopping ($\kappa_{12}/{\rm 2}\pi$ = 2 kHz) is shown in Fig.\,4(a). The parameter $2g_{r}/\kappa_{12}=$25 is used in Fig.\,4(b). In the simulation, a 2$\pi$ red-sideband pulse is applied to the second ion at $t=\pi/4\kappa_{12}=62.5$ $\mu s$.

In principle, our scheme is not limited by the number of ions. However, due to the phonon-number dependence on the sideband Rabi frequency, our scheme does not work in the presence of multiple local phonons in a trapped-ion chain. We discuss the scalability of phase shift operations in terms of the number of ions and that of phonons in Sec. IV.

\section{Experimental results}

\subsection{Experimental setup}

We perform experiments with two $^{40}{\rm Ca}^{+}$ ions trapped in a linear Paul trap. The frequencies for harmonic confinement along the radial ($\it{x}$ and $\it{y}$) and axial ($\it{z}$) directions for two ions are $(\omega_{x}, \omega_{y}, \omega_{z})/$2$\pi$=(3.0, 2.8, 0.11) MHz, where an RF voltage is stabilized using a method similar to that given in \cite{41}. The internal states $\ket{S_{1/2},m_{j}=-1/2}\equiv\ket{\downarrow}$ and $\ket{D_{5/2},m_{j}=-1/2}\equiv\ket{\uparrow}$ are used to encode the two-level system.

Each experiment starts with Doppler cooling using 397 nm (${\it S_{\rm 1/2}}$--${\it P_{\rm 1/2}}$) and 866 nm (${\it D_{\rm 3/2}}$--${\it P_{\rm 1/2}}$) lasers. Then, ground state cooling of the radial motional modes ($x$ and $y$) is carried out using resolved sideband cooling with a 729 nm laser (${\it S_{\rm 1/2}}$--${\it D_{\rm 5/2}}$). In the present experiment, we employ the local phonon mode along the $y$ direction, and the average motional number for the $y$ direction is 0.04.

\subsection{Dynamical decoupling of a single local phonon}

\begin{figure}[t]
\centering
  \includegraphics[width=8.5cm]{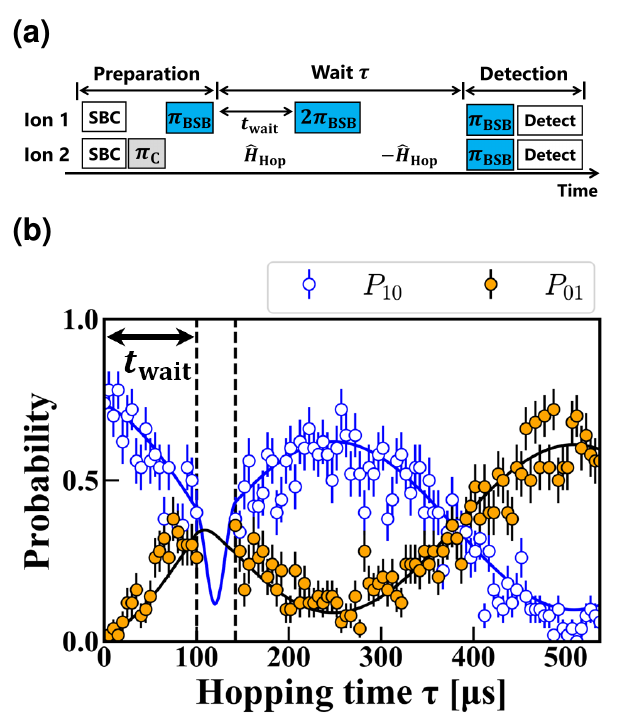}
\caption{\label{fig2} (a) Experimental sequence. After sideband cooling (SBC), the phonon state $\ket{1,0}$ is prepared by applying a carrier $\pi$ pulse ($\pi_{\rm C}$) to Ion 2 and a blue-sideband (BSB) $\pi$ pulse to Ion 1. The hopping time $\tau$ is varied to observe the phonon dynamics. We wait for a duration $t_{\rm wait}=100\,\mu$s, and then a 2$\pi$ BSB pulse is applied to Ion 1. To perform fluorescence detection (Detect), BSB $\pi$ pulses are applied to both ions to map the probability amplitude onto the internal state. (b) Results of the DD with a single sideband pulse. The solid curves are numerically calculated results. }
\end{figure}

\begin{figure*}[t]
\centering
  \includegraphics[width=15.0cm]{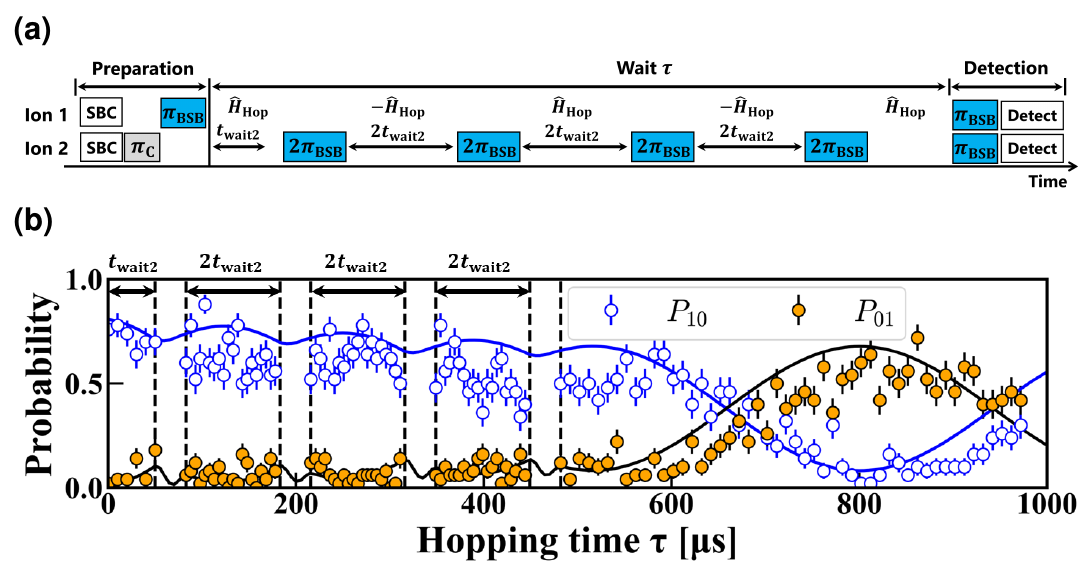}
\caption{\label{fig1} (a) Experimental sequence for controlling a local phonon with multiple sideband pulses. After preparing the ions in $\ket{1,0}$, the sequence of 2$\pi$ pulses at the blue-sideband is applied to Ion 2. The duration between each BSB 2$\pi$ pulse is set to be 2$t_{\rm wait2}$. During the hopping period, four 2$\pi$ BSB pulses are applied. Finally, a mapping pulse is applied, followed by the detection. (b) Results of the DD technique with multiple sideband pulses. The dashed curves are numerically calculated results. Each data point is an average of 50 measurements.}
\end{figure*}

We implement single-local-phonon control with a single blue-sideband 2$\pi$ pulse for two ions. The experimental sequence is given in Fig.\,5(a). After the sideband cooling, we prepare the ions in the state $\ket{\psi_{\rm Init}}=\ket{\psi_{1}}\otimes\ket{\psi_{2}}=\ket{\uparrow_{1},1}\otimes\ket{\uparrow_{2},0}\equiv\ket{1,0}$. This state is generated by applying a carrier $\pi$ pulse to Ion 2, which is followed by a blue-sideband $\pi$ pulse to Ion 1. The hopping time $\tau$ is varied to observe the phonon dynamics. After preparing the ions in $\ket{1,0}$, a fixed wait time $t_{\rm wait}=100\,\mu$s is applied, and then a blue-sideband 2$\pi$ pulse is applied to Ion 1.  The time evolution of the phonon state after this pulse in the transformed basis is governed by the transformed Hamiltonian $-\hat{H}_{\rm Hop}$. Note that this scheme also works in the case of applying a 2$\pi$ sideband pulse to Ion 2, as seen earlier. 

After waiting for $\tau$, a blue-sideband $\pi$ pulse is applied to both ions to map the probability amplitude of $\ket{\uparrow_{i},1}$ onto $\ket{\downarrow_{i},0}$. The ions are then illuminated with lasers at 397 nm and 866 nm to collect the state-dependent fluorescence with a CCD camera. We count the events when one ion fluoresces while the other does not, to calculate the probability of detecting $\ket{1,0}$ ($P_{10}$) [$\ket{0,1}$ ($P_{01}$)].

The results are shown in Figs.\,5(b). The blue and orange data are the probability of detecting the state $\ket{1,0}$ and $\ket{0,1}$, respectively. Each data point is an average of 50 measurements. The time step for each data is $5\,\mu$s. The solid curves in Fig. 5(b) are numerically calculated results using the Lindblad master equation with the parameter $\kappa_{12}/{\rm 2}\pi$ = 1.9 kHz. The parameter $\kappa_{12}$ is obtained by fitting the free hopping result with a sinusoidal function. The imperfection of state preparation and the dephasing of carrier and blue-sideband transitions are also included in the numerical calculation. The experimental data show agreement with the numerically calculated results.

We also demonstrate DD using multiple sideband pulses. The experimental sequence is shown in Fig.\,6(a). In this experiment, we apply multiple blue-sideband 2$\pi$ pulses to Ion 2 so that the phonon localizes to a particular ion site. After preparing the ions in $\ket{\psi_{\rm Init}}=\ket{1,0}$, we wait for a duration of $t_{\rm wait2}=50\,\mu$s and then apply a blue-sideband 2$\pi$ pulse to Ion 1. In the experiment, we sequentially apply a blue-sideband $2\pi$ pulse so that the phonon is localized to Ion 1.

The result is shown in Fig.\,6(b). The blue and orange data are the probability of detecting $\ket{1,0}$ and $\ket{0,1}$, respectively. Each data point is an average of 50 measurements. Here, a hopping rate of $\kappa_{12}/$2$\pi\approx1.76$ kHz is used, which is obtained by fitting the free hopping result with a sinusoidal function. The data points shown between $\tau=0$ and $\tau=t_{\rm wait2}=50\,\mu$s and the free hopping part at the last (440--1000\ $\mu$s) are collected with a time step of $10\,\mu$s, while other data points shown in Fig.\,6(b) are collected with a time step of $5\,\mu$s. The dashed curves are numerically calculated results using the Lindblad master equation as in the former experiment. As seen in Fig.\,6(b), a single phonon stays in Ion 1, and the experimental data show agreement with the numerically calculated results.

\section{Discussion}

The blue-sideband $\pi$ pulse infidelity ($\sim0.08$) limits the contrast of the experimental data. We assume that the ac Stark shifts may cause this infidelity due to the off-resonant coupling to the carrier transition of the sideband transition \cite{42}. We find that the 729 nm beam illuminating one of the ions interferes with the tail of another beam directed at the other ion around the plane in which the axis for the ion string resides. We speculate that this beam interference causes fluctuations in the beam intensity experienced by the ions, resulting in a variation of ac Stark shifts for the relevant transitions. Therefore, an ac Stark compensator may improve the overall contrast of the data.

In the following, we discuss the scalability of the presented method in terms of (1) the number of ions and (2) the number of phonons. The former can be judged with respect to the minimal distance between the ions. Whether the method presented here is applicable depends on the ratio between the speed of manipulating local phonons and each hopping rate, where the hopping rate scales in proportion to the inverse of the cube of the inter-ion distance \cite{15}. Even if the number of ions in the trap is increased, as long as the minimum distance between the ions is longer than a certain value, the method is applicable. It should be noted that this requires increasingly shallow axial potentials for large numbers of ions in a single linear trap. By using an array of independent single-ion traps and local phonons in a system \cite{12,13}, we can avoid the use of such shallow potentials and increase the number of ions. 

As for the scalability in terms of the number of phonons, as noted earlier, operations on multiple local phonons cannot be performed with resonant sideband pulses and may require dispersive shifts, as assumed in the scheme proposed in Ref.\, [25]. For future work, we discuss the experimental implementation of the phase-shift operator using practical parameters. As described above, the condition $\chi\gg\kappa_{ij}$ must be satisfied for the phase-shift operation. This condition can be realized by (1) increasing the sideband Rabi frequency $2g_r$, (2) decreasing the detuning $\Delta$, or (3) decreasing the hopping rate $\kappa_{ij}$. 

Increasing the sideband Rabi frequency is a straightforward way to enhance the dispersive interaction strength. The sideband Rabi frequency is practically limited by the off-resonant excitation of the carrier transition \cite{43}, and the typical value is less than 100 kHz. Likewise, the detuning $\Delta$ is also determined by the off-resonant excitation of the sideband transition. When an ion is irradiated with an off-resonant red-sideband pulse with $\Delta$ detuning from the sideband transition, the off-resonant excitation probability is $P_{\Delta} \sim 4{g_r}^2/\Delta^2$, where $2g_r$ is the Rabi frequency at the red-sideband transition. This relation indicates that to suppress the excitation error below 1\%, the detuning $\Delta$ needs to be larger than $10\times 2g_r$. Then, we get $\chi=g_{r}^2/\Delta=g_{r}^2/(10\times 2g_r)=2g_r/40=2\pi\times2.5$ kHz, where $2g_r$ is assumed to be $2\pi\times100$ kHz. Therefore, in practice, we need to decrease the hopping rate to realize $\chi\gg\kappa_{ij}$.

\section{Conclusions}
In conclusion, we have demonstrated DD of local phonon modes in a two-ion chain. A 2$\pi$ pulse at the blue-sideband transition induces a sign flip of a single-local-phonon state, reversing the dynamics of the local phonon. Our work provides a new tool for engineering local phonon couplings. 

\section*{Acknowledgments}
The authors wish to thank Cl\'{e}ment Chamboulive for his contribution in the early stage of this work. This work was supported by MEXT Quantum Leap Flagship Program (MEXT Q-LEAP) Grant Number JPMXS0118067477. R. O. was supported by JSPS KAKENHI Grant Number JP21J10054.

\end{document}